\documentclass[final,english]{bullsrsl}

\usepackage[latin1]{inputenc}

\usepackage[T1]{fontenc}
\usepackage{float}

\usepackage{natbib} 

\usepackage{graphicx}

\begin{document}
\title{Detection and Properties of Lyman Break Galaxies at 1.6 $\leq$ z $\leq$ 3.0 in GOODS-North}



  \author[affil={1}, corresponding]{Bibhuprasad} {Mishra}
  \author[affil={2}]{Kanak} {Saha}
  \author[affil={3}]{Divya} {Pandey}
  \author[affil={1}]{Ananta C.} {Pradhan}

\affiliation[1]{National Institute of Technology, Rourkela, Odisha, 769008, India}
\affiliation[2]{Inter-University Centre for Astronomy and Astrophysics, Pune,  Maharashtra, 411007, India}
\affiliation[3]{Tartu Observatory, University of Tartu, Observatooriumi 1, T\~oravere, 61602, Estonia}


\correspondance{bibhuprasad2203@gmail.com}


\maketitle

\begin{abstract}
Lyman Break Galaxies (LBGs) are actively star-forming galaxies identified through their characteristic rest-frame ultraviolet spectral break, making them powerful probes of galaxy evolution across cosmic time.  We present a search for LBGs at redshift $1.6 \leq z \leq 3.0$ in the GOODS-North field using the dropout technique combined with color selection from 3DHST data. We utilize two selection criteria to identify LBG candidates at $z \sim 2.1$ and $z \sim 2.7$ via F275W- and F336W-dropouts, yielding 192 and 135 newly identified LBG candidates in the redshift ranges $1.6 \leq z \leq 2.3$ and $2.3 < z \leq 3.0$, respectively. This study presents an estimation of the physical properties, including age, mass, and SFRs, of the LBG candidates using the spectral energy distribution (SED) fitting method. The morphological analysis shows that a majority of galaxies in our sample exhibit disc-like structure. Overall, the results demonstrate that LBGs are low-to intermediate-mass compact structures with effective half-light radii of approximately 3.5 kpc.
\end{abstract}





\section{Introduction}
Lyman Break Galaxies (LBGs) represent one of the important classes of high-redshift galaxies for understanding galaxy evolution at early cosmic epochs.  First identified by \citet{Steidel1993}, LBGs are typically compact, star-forming systems dominated by relatively young stellar populations. They are observed to have strong ultraviolet (UV) emission, pointing to vigorous ongoing star formation. They are characterized by strong rest-frame UV emission, indicative of intense, ongoing star formation, and are detected through the Lyman Break dropout technique, which has enabled the detection of thousands of LBG candidates at redshifts z $\sim$ 3-8 using ground-based observations \citep{Steidel1995}. However, only a few major studies have been done at redshifts around z $\sim$ 1-3 due to the lack of highly sensitive space-based telescopes to observe near-UV (NUV). Recent NUV observations from various surveys (e.g., GALEX, HST, and UVIT)  enable us to adapt the dropout selection criteria to detect LBG candidates at z $\sim$2. The properties of high-redshift LBG candidates are difficult to study because their rest-frame optical wavelengths fall beyond the observational limit. In contrast, the low-redshift LBGs offer significant advantages, allowing us to investigate them in both rest-frame UV and optical wavelengths. Hence, it is important to gain insight into the low-redshift LBGs to understand the physical and morphological nature of the high-redshift ones. We aim to detect and study LBGs in the redshift range 1.6 $\leq$ z $\leq$ 3.0 because this epoch corresponds to the peak of the cosmic star formation history, often referred to as cosmic noon \citep{Madau2014}. In this redshift window, galaxies are rapidly building up their stellar mass, undergoing intense star formation, and experiencing significant morphological transformations such as gas accretion, feedback, and mergers (\cite{ForsterSchreiber2020}) that shape galaxy evolution and set the stage for the mature galaxy population observed at lower redshift. Throughout this work, we adopt a flat $\Lambda$CDM cosmology with $H_0 = 70\,\mathrm{km\,s^{-1}\,Mpc^{-1}}$, $\Omega_m = 0.3$, and $\Omega_\Lambda = 0.7$.

\section{Observation and Analysis}

The GOODS-North field \citep{Giavalisco2004}, centered around the Hubble Deep Field North (HDFN; \cite{Williams1996}) at coordinates  RA = $189.2058^{\circ}$, DEC = $+62.2161^{\circ}$, covers an area of  $\sim$171 arcmin$^2$. This region has been the target of some of the deepest observations ever conducted with HST, Spitzer, and other world-class telescopes. The 3D-HST survey of GOODS-N \citep{Brammer2012} provides insights into the physical processes that drive the evolution of distant galaxies. It includes two primary WFC3/G141 grism observations, enabling the critical third-dimension spectroscopic redshifts for galaxies at z $>$ 1.  For the present analysis, photometric data are obtained from the 3D-HST catalog, which consists of various observations including HST (WFC3 and ACS), Subaru and KPNO optical imaging, 2MASS near-infrared (NIR) observations, NOAO/Ks, LRIS, and Spitzer/IRAC mid-infrared channels. The NUV photometry used in this work was obtained from the Hubble Deep UV Legacy Survey (HDUV; \citep{Oesch2018}), which provides deep HST/WFC3-UVIS imaging in the F275W and F336W bands over the GOODS-North field. These NUV measurements were combined with the multiwavelength photometric catalogs from the 3D-HST survey (see \cite{Skelton2014} for details of the filters) for sample selection and analysis. The combined coverage provides a continuous spectral baseline from the UV to the mid-IR. This multi-band photometry enables the application of color-selection techniques to identify LBG candidates. To select LBG candidates in the redshift range 1.6 $\le z \le$3.0, we employ a color-based dropout technique, which identifies galaxies exhibiting a characteristic suppression of UV flux. This method exploits the sharp spectral break at the Lyman limit (912~\AA\ in the rest frame). For galaxies at 1.6 $\leq$ z $\leq$2.3, the break is redshifted into the HST/WFC3 F275W band, resulting in F275W dropouts, while galaxies at 2.3$ <$ z $\leq$3.0 are identified as F336W dropouts. Applying the F275W and F336W dropout selection criteria (see \citealt{Hathi2010} for details), we identify 192 and 135 LBG candidates in the redshift ranges 1.6 $\leq$ z $\leq$2.3 and 2.3 $<$ z $\leq$3.0, respectively. To the best of our knowledge, no previous study has presented a dedicated catalog of F275W- and F336W-dropout selected LBG candidates in GOODS-North over this redshift range. 

\begin{figure}[H]
\centering

\begin{minipage}{0.49\linewidth}
    \centering
    \includegraphics[width=\linewidth]{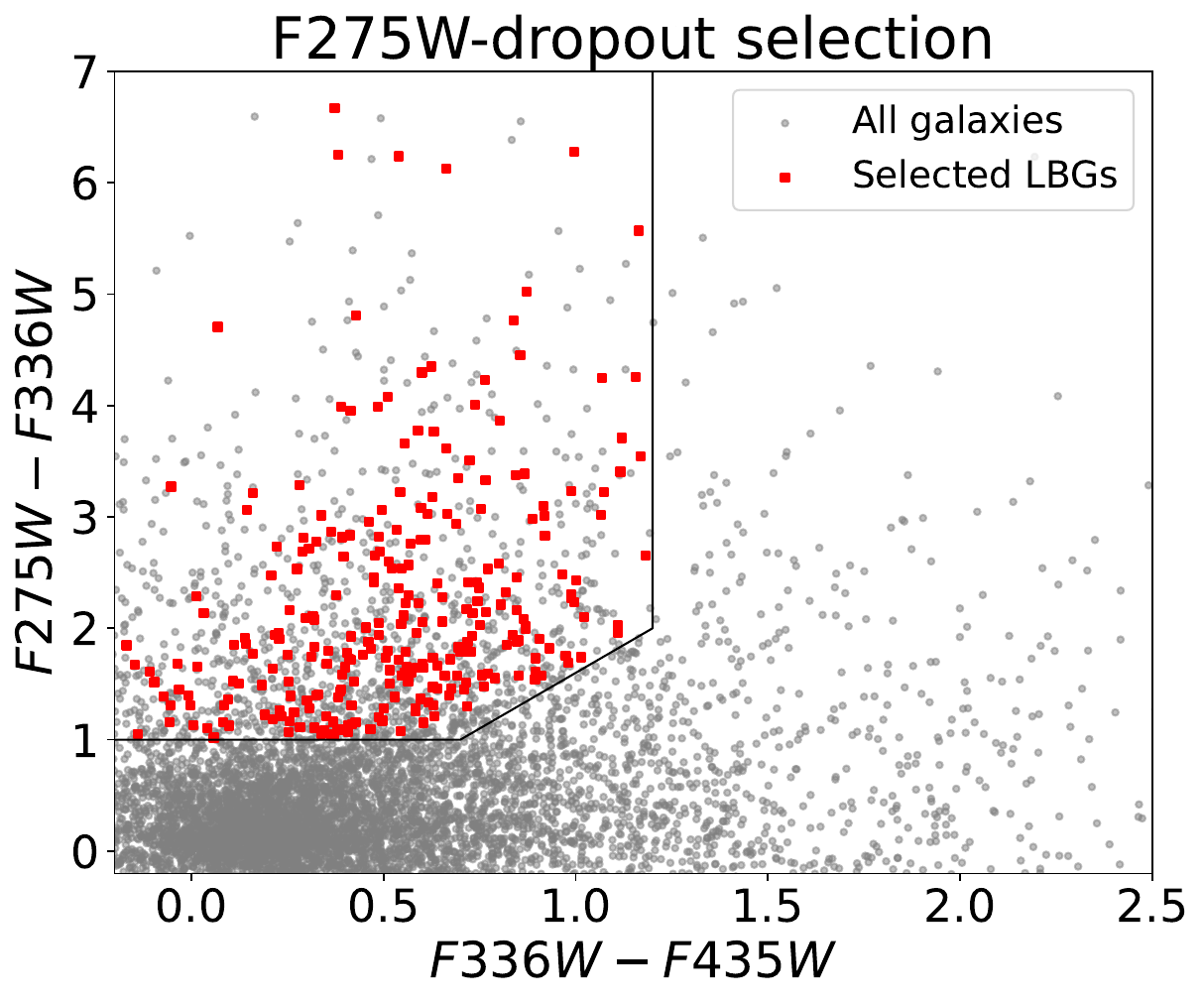}
\end{minipage}\hfill
\begin{minipage}{0.49\linewidth}
    \centering
    \includegraphics[width=\linewidth]{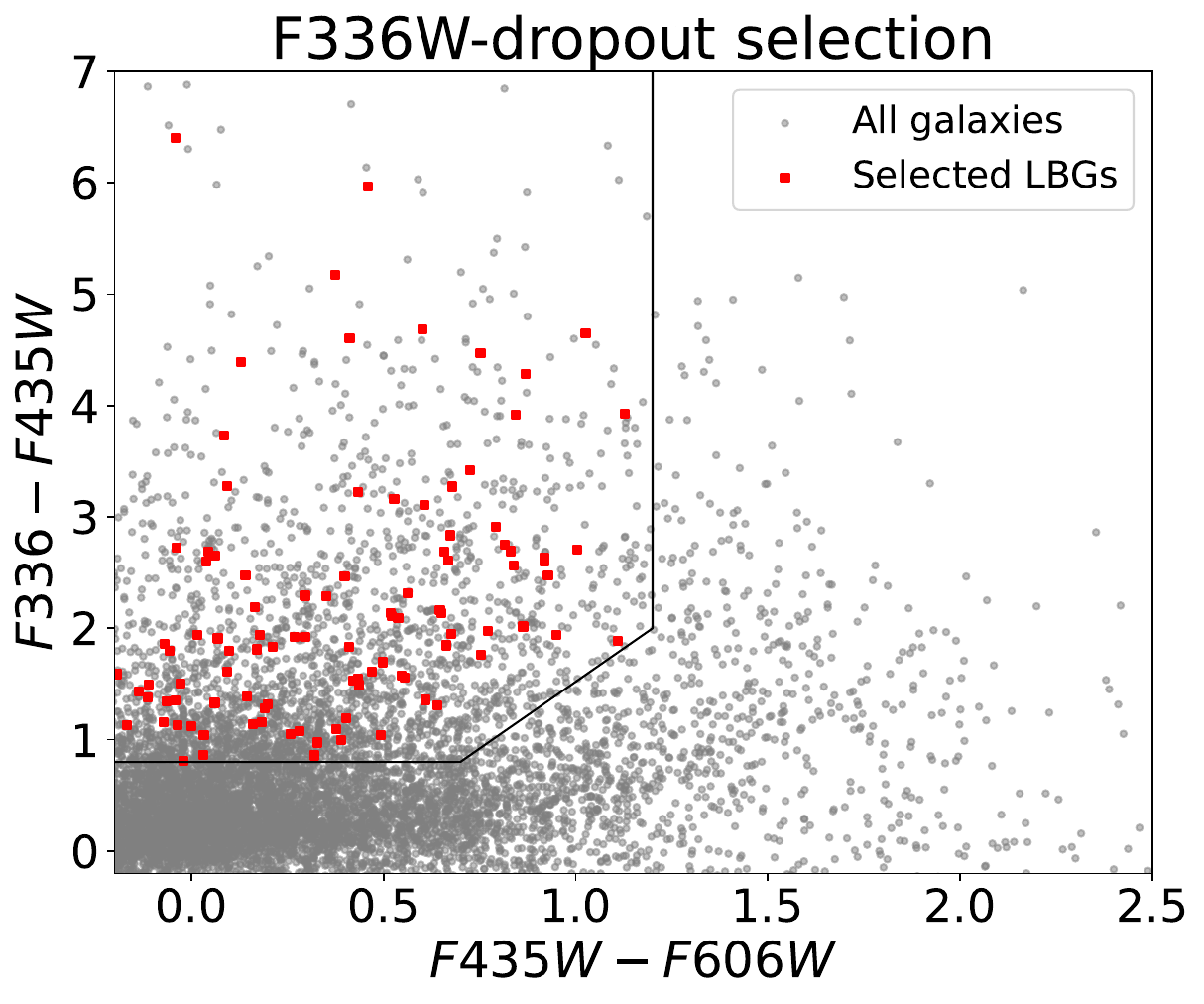}
\end{minipage}

\begin{minipage}{12cm}
\caption{The left panel shows the color-color diagram with the F275W-dropout selection region, while the right panel shows the F336W-dropout selection region. Gray points represent all objects in the catalog. Red points indicate the selected LBG candidates, whereas gray points within the selection region were excluded based on the additional selection criteria.}
\label{fig:1}
\end{minipage}

\end{figure}

\begin{figure}[H]
\centering

\includegraphics[width=\linewidth]{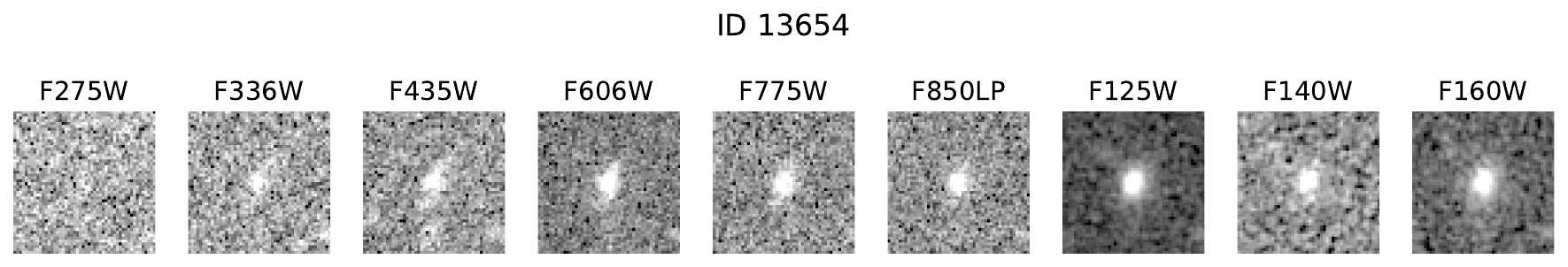}

\includegraphics[width=\linewidth]{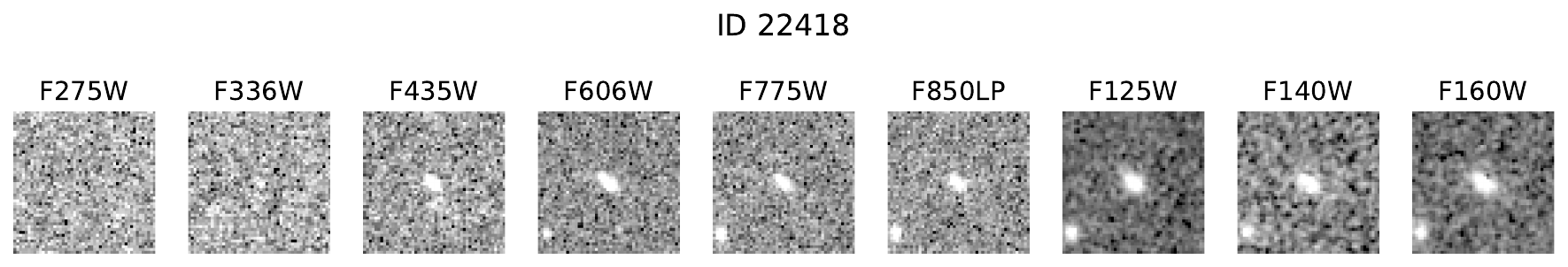}

\begin{minipage}{12cm}
\caption{Example of two LBG candidates with $50\times50$ pixel cutouts. The top panel shows an F275W-dropout galaxy (ID 13654) at redshift $z=1.77$, and the bottom panel shows an F336W-dropout galaxy (ID 22418) at $z=2.38$.The nine displayed HST images are shown for visualization only; the SED fitting was performed using the full multiwavelength photometric catalog, including HDUV, optical/NIR HST, ground-based, and Spitzer/IRAC data.}
\label{fig:2}
\end{minipage}

\end{figure}

We therefore consider these objects to be newly identified as LBG candidates through the present selection technique. In Figure~\ref{fig:1}, the gray data points represent all the sources from the 3DHST catalog that are not selected as LBG candidates, while the red points indicate the selected Lyman-dropout candidates. Sources within the selection regions were excluded if they did not satisfy all selection criteria samples, mainly because they failed to meet all the defined criteria. Figure~\ref{fig:2} shows two representative examples of our LBG candidates: ID 13654 (F275W-dropout) and ID 22418 (F336W-dropout), which are undetected in their respective dropout bands but are clearly detected in redder filters. 

\section{Results and Discussion}

\begin{figure}[H]
\centering

\begin{minipage}{0.49\linewidth}
    \centering
    \includegraphics[width=\linewidth]{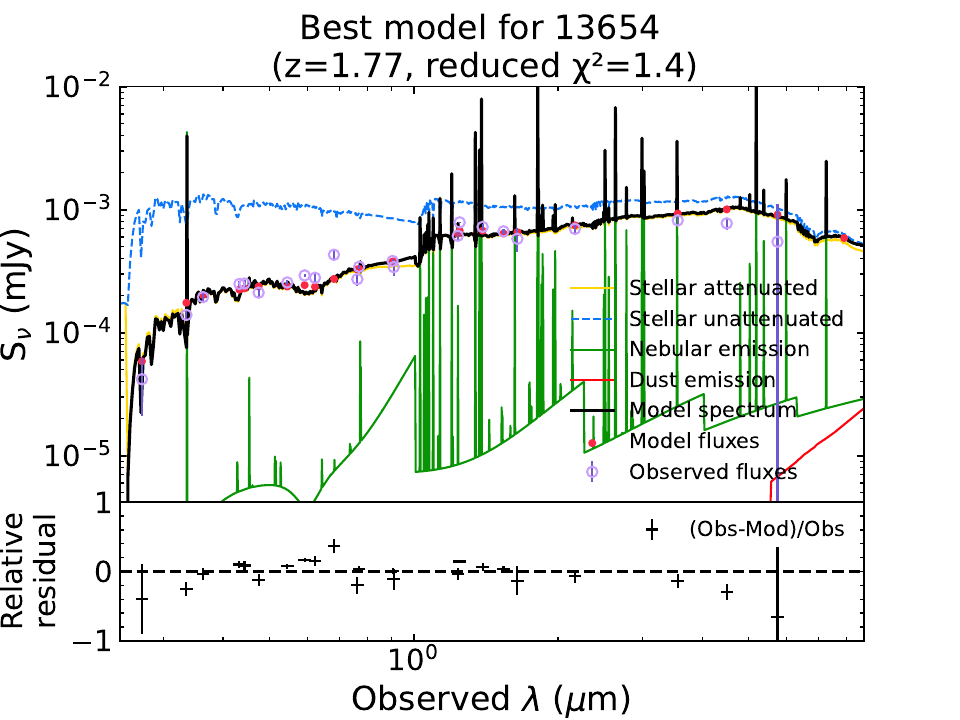}
\end{minipage}\hfill
\begin{minipage}{0.49\linewidth}
    \centering
    \includegraphics[width=\linewidth]{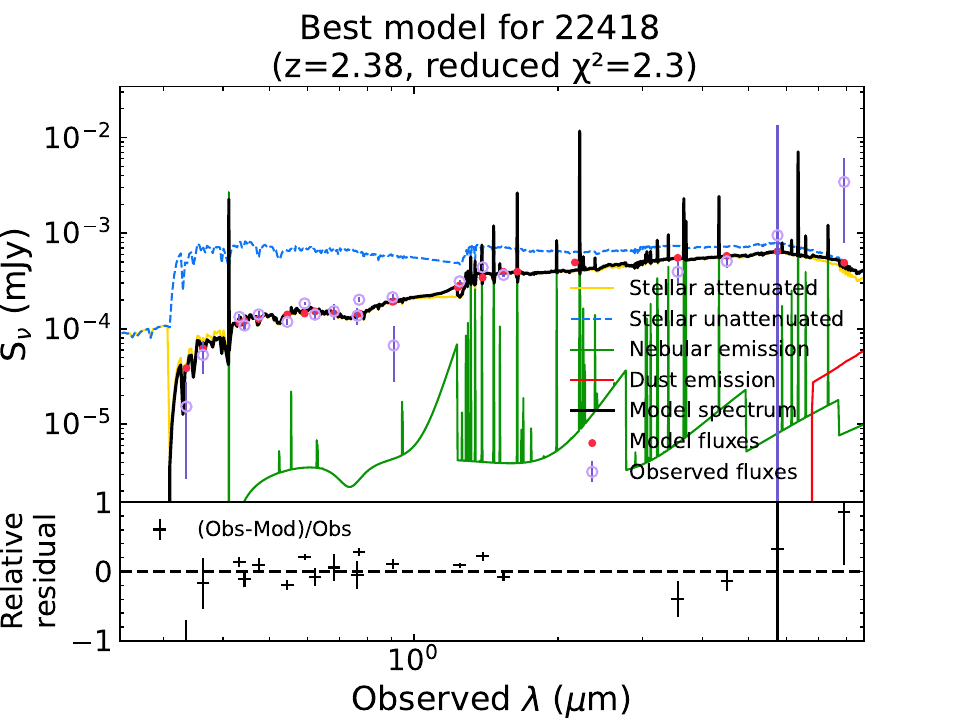}
\end{minipage}

\vspace{2mm}

\begin{minipage}{12cm}
\caption{Spectral energy distribution (SED) fits for two representative LBG candidates obtained using CIGALE. The observed multiwavelength photometric data are shown together with the best-fitting model SEDs.}
\label{fig:3}
\end{minipage}

\end{figure}

Spectral energy distributions (SEDs) spanning a broad wavelength range provide robust constraints on the physical properties of galaxies. We model the SEDs using the widely adopted code CIGALE \citep{Boquien2019}, compiling photometric measurements for the candidates across near-UV, optical, and NIR bands. We use a double-exponential star formation history (sfh2exp), \cite{Chabrier2003}, an initial mass function (IMF), \cite{Bruzual2003}, stellar population synthesis models (bc03), a power-law dust attenuation prescription (dustatt\_powerlaw), and infrared dust emission templates from \cite{Dale2014}. We adopt spectroscopic redshifts when available and use photometric redshifts for the remaining sources as the input to the SED modeling. We determined the best-fit SEDs through reduced $\chi^2$ minimization. Two example SEDs are shown in Figure~\ref{fig:3} for sources ID 13654 and 22418. For the morphological analysis, we use StatMorph (\cite{RodriguezGomez2019}), a Python package that provides non-parametric (e.g., Gini, M20) and parametric (e.g., S\'ersic).We measured the Gini and M20 parameters from the HST/WFC3 F160W imaging, with the corresponding F160W PSF supplied during the analysis. We adopt the elliptical half-light radius "rhalf\_ellip" as the effective radius, $R_e$. Although the HST imaging provides sufficient resolution to resolve the overall galaxy structure, non-parametric morphology measurements at $z\sim$2-3 may still be influenced by finite spatial resolution and S/N limitations. Therefore, the resulting classifications should be interpreted as a broad characterization of the sample.

We find that the stellar masses of the LBG sample span the range $10^8$-$10^{11}\, M_\odot$, with a trend of increasing stellar mass with redshift. for the particular dropout selection set, and the mean SFRs over the last 100 Myr are 14.7 $M_\odot$ $yr^{-1}$ and 20.9 $M_\odot$ $yr^{-1}$ for the redshift ranges of 1.6 $\leq$ z $\leq$ 2.3 and 2.3 $<$ z $\leq$ 3.0, respectively. For comparison, we constructed a sample of 200 galaxies from the 3D-HST catalog within the same redshift range as the LBG sample. Objects satisfying the adopted F275W and F336W dropout-selection criteria were excluded from this sample. The resulting non-LBG sample was used to compare the general properties of galaxies selected and not selected by the dropout technique. The left panel of Figure 4 shows the stellar mass distributed over the redshift range with SFR as the color map; the right panel shows the comparison, indicating that LBG candidates tend to exhibit higher SFRs than the non-LBG sample, which shows mean values of 2.3 $M_\odot$ $yr^{-1}$ and 5 $M_\odot$ $yr^{-1}$ for the respective redshift range. This difference is expected, at least in part, because the dropout-selection method preferentially identifies actively star-forming galaxies with strong UV emission; in both figures, the stellar mass is in $\log_{10}(M_\ast/M_\odot)$ units. A fraction of these galaxies exhibit much higher activity, with SFRs reaching up to $\sim$100 $M_\odot$ $yr^{-1}$, highlighting their role as vigorously star-forming systems in the early universe. A clear trend is observed: galaxies with higher stellar masses exhibit higher SFRs. After inspection, we find the median values of the Gini and M20 coefficients to be 0.49 and -1.49, respectively (see \cite{Lotz2004} for the morphological classification). Morphological classification was performed using the Gini-M20 plane. Among the 327 LBG candidates, 216 ($\sim66\%$) are classified as disk-like, 21 as elliptical, and 70 as irregular, while the remaining 20 are unclassified. The full Gini-M20 distribution and classification regions are shown in Figure 5. The LBG candidates have effective radii of approximately 3.4-3.5 kpc, corresponding to characteristic diameters of 6.7-7.0 kpc. Given the small difference in the measured sizes, we do not infer a significant size evolution over the redshift range explored in this study.

\begin{figure}[ht]
\centering

\begin{minipage}{0.49\linewidth}
    \centering
    \includegraphics[width=\linewidth]{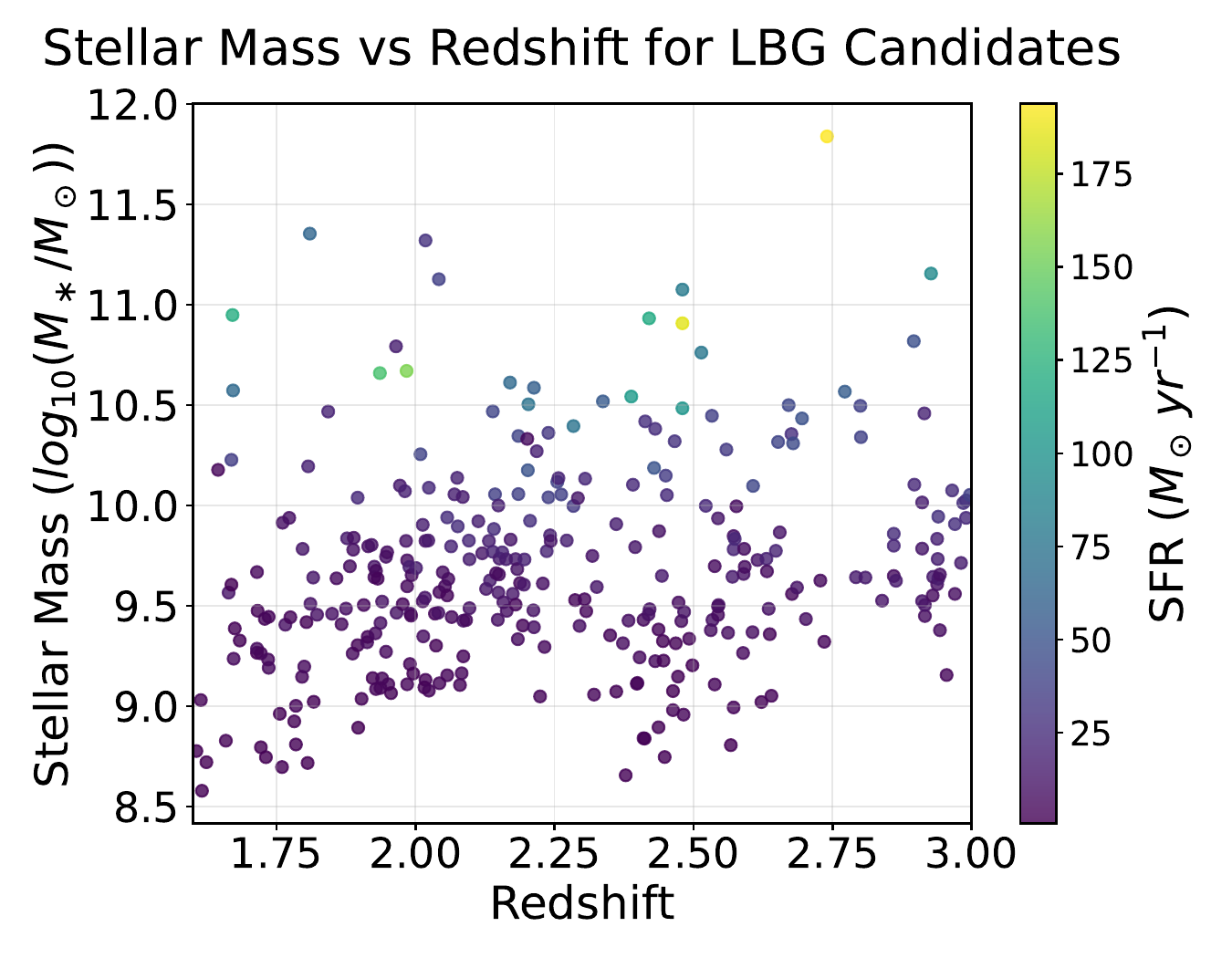}
\end{minipage}\hfill
\begin{minipage}{0.49\linewidth}
    \centering
    \includegraphics[width=\linewidth]{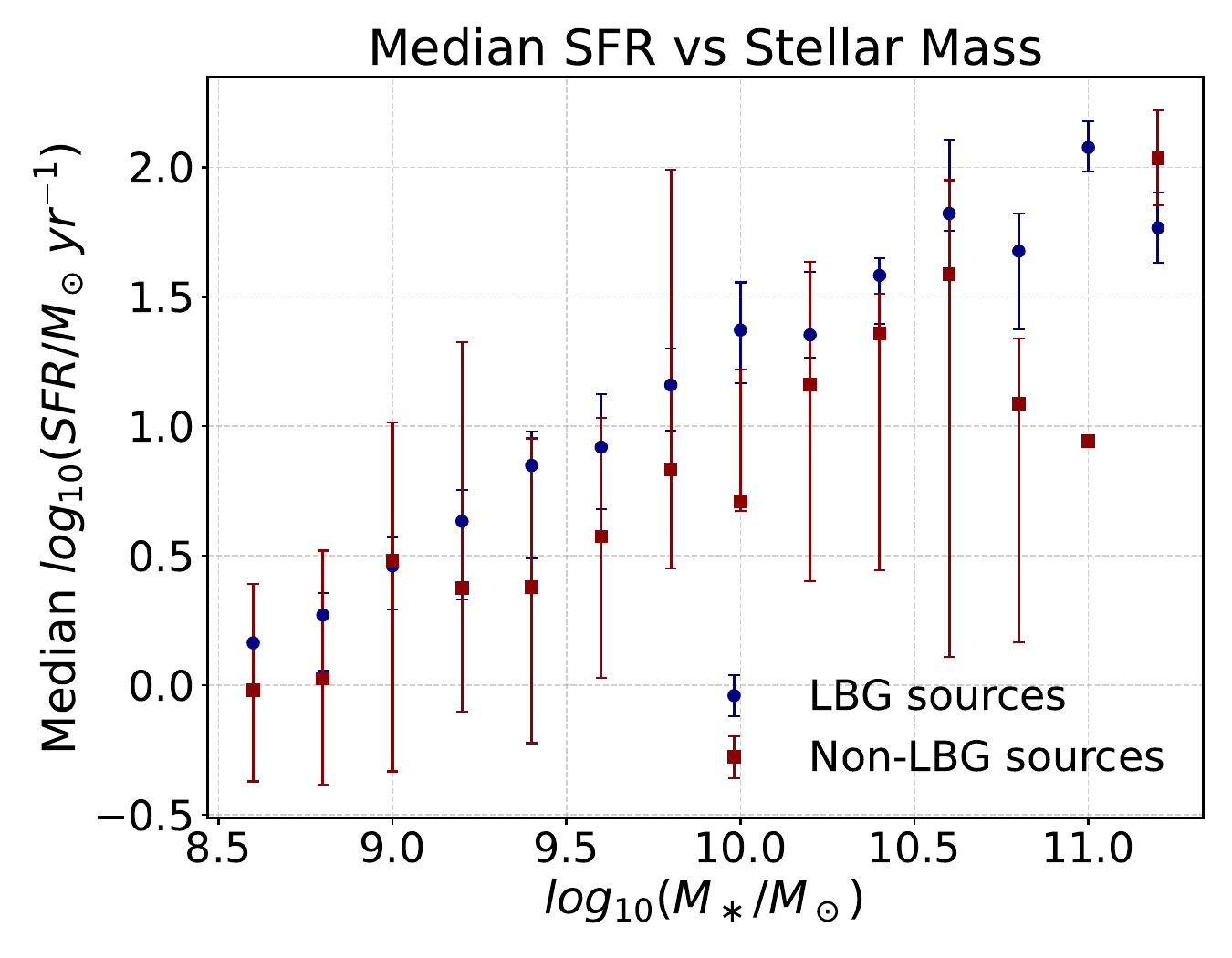}
\end{minipage}

\vspace{2mm}

\begin{minipage}{12cm}
\caption{Left panel: Stellar mass as a function of redshift, with the star formation rate (SFR) represented by the color scale. Right panel: Median SFR as a function of stellar mass for the LBG and non-LBG samples. The error bars represent the 16th and 84th percentiles. Stellar mass is expressed as $\log_{10}(M_\ast/M_\odot)$ and SFR in $M_\odot\,\mathrm{yr}^{-1}$.}
\label{fig:4}
\end{minipage}

\end{figure}

\begin{figure}[!t]
\centering

\includegraphics[width=\columnwidth]{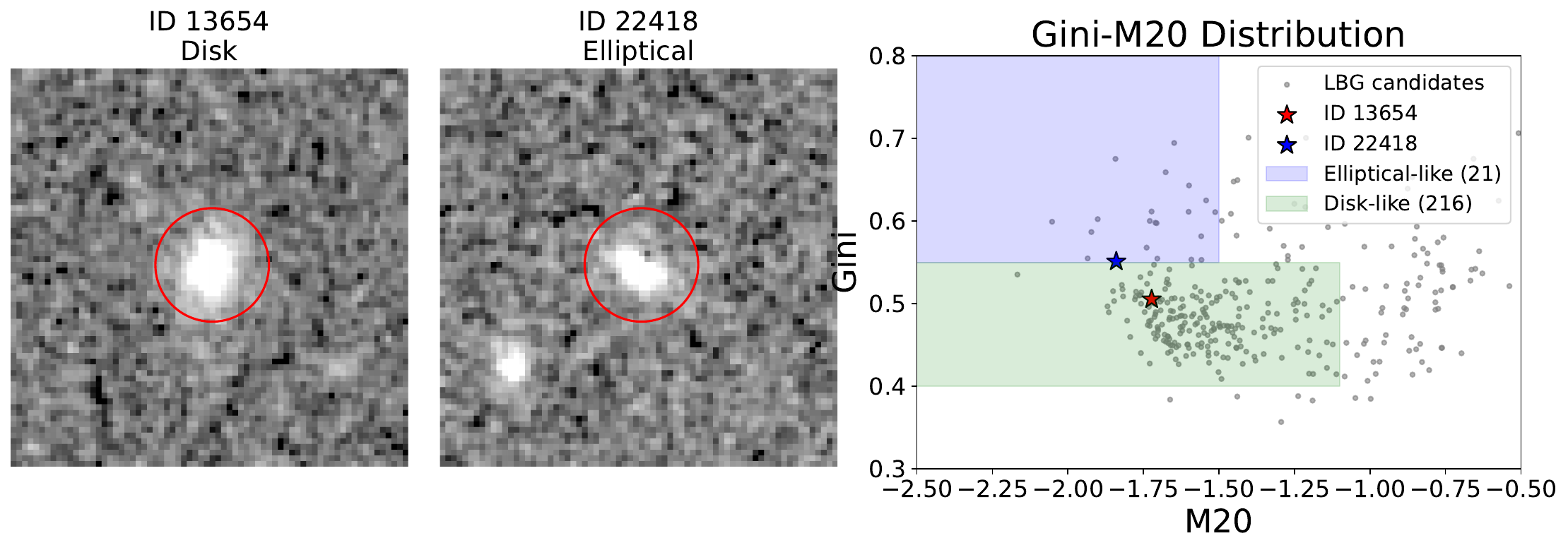}

\vspace{2mm}

\begin{minipage}{12cm}
\caption{Distribution of the 327 LBG candidates in the Gini--$M_{20}$ plane. The shaded regions indicate the adopted morphological classification boundaries for disk-like and elliptical galaxies. Galaxies outside these regions are classified as irregular or remain unclassified according to the adopted criteria.}
\label{fig:GM20}
\end{minipage}

\end{figure}

In this work, we report a sample of 327 newly identified LBG candidates in the redshift range 1.6 $\leq$ z $\leq$3 using the dropout technique and conduct detailed SED modeling. These galaxies predominantly fall in the low- to intermediate-mass range and exhibit relatively high SFRs, consistent with their classification as actively star-forming systems. A morphological analysis further reveals that the majority of these galaxies exhibit disk-like structures, typically spanning only a few kpc in diameter. Our results shed light on the fundamental properties of LBGs and underscore their significance in tracing the processes of galaxy formation and evolution at cosmic noon.


\begin{acknowledgments}
This work makes use of imaging and catalog data from the Hubble Deep UV Legacy Survey (HDUV; Oesch et al. 2018) and the 3D-HST Treasury program (Brammer et al. 2012; Skelton et al. 2014), obtained with the NASA/ESA Hubble Space Telescope and retrieved from the Mikulski Archive for Space Telescopes (MAST).
\end{acknowledgments}

\begin{furtherinformation}

\begin{orcids}

  \orcid{0009-0008-7388-0957}{Bibhuprasad}{Mishra}
  \orcid{0000-0002-8768-9298}{Kanak}{Saha}
  \orcid{0000-0002-4656-056X}{Divya}{Pandey}
  \orcid{0000-0001-5808-6001}{Ananta C.}{Pradhan}

\end{orcids}

\begin{authorcontributions}
This work is the result of a long-term collaboration to which all authors have made significant contributions.
\end{authorcontributions}

\begin{conflictsofinterest}
 The authors declare no conflict of interest
\end{conflictsofinterest}

\end{furtherinformation}



%

\bibliographystyle{bullsrsl-en}

\bibliography{extra.bib}

\end{document}